\documentstyle[12pt,epsfig]{article}
\columnwidth\textwidth

\begin{document}
\renewcommand{\theequation}{\arabic{equation}}
\parskip=4pt plus 1pt
\textheight=8.7in
\textwidth=6.0in
\newcommand{\be}{\begin{equation}}
\newcommand{\ee}{\end{equation}}
\newcommand{\bea}{\begin{eqnarray}}
\newcommand{\eea}{\end{eqnarray}}
\newcommand{\co}{\; \; ,}
\newcommand{\nn}{\nonumber \\}
\newcommand{\scs}{\co \;}
\newcommand{\per}{ \; .}
\newcommand{\unith}{{\bf{\mbox{1}}}}
\newcommand{\la}{\langle}
\newcommand{\ra}{\rangle}
\thispagestyle{empty}
\begin{flushright}
NORDITA-95/77 N,P\\
BUTP-95-34\\
UWThPh-1995-34\\
HU-TFT-95-64
\end{flushright}

\vspace{2cm}
\begin{center}
\begin{Large}
Elastic $\pi\pi$ scattering to two loops
\\[1cm] \end{Large}
J. Bijnens$^1$, G. Colangelo$^{2}$, G. Ecker$^3$, J. Gasser$^2$ and M.E.
Sainio$^4$ \\[2cm]
${}^1$ NORDITA, Blegdamsvej 17, DK--2100 Copenhagen\\
${}^2$ Inst. Theor. Phys., Univ. Bern, Sidlerstr. 5, CH--3012 Bern\\
${}^3$ Inst. Theor. Phys., Univ. Wien, Boltzmanng. 5, A--1090 Wien\\
${}^4$ Dept. of Physics, Univ. Helsinki, P.O. Box 9, FIN--00014 Helsinki
\\[1cm]
November 1995\\[1cm]
\end{center}
\begin{abstract}
\noindent
We evaluate analytically the elastic $\pi\pi$ scattering amplitude to two
loops in chiral perturbation theory and give numerical values for the two
$S$--wave  scattering lengths and for the phase shift difference
$\delta_0^0-\delta_1^1$.
\end{abstract}

\clearpage
\noindent
1. In the framework of chiral perturbation theory (CHPT)
\cite{wein79,glann,hlann},
the elastic $\pi\pi$ scattering amplitude is evaluated by
 an expansion
 in powers of the external momenta and of the light quark masses,
\bea
A=A_2+A_4+A_6+\cdots\co
\label{expansiona}
\eea
where $A_n$ is of  $O(p^n)$. The first two terms in this expansion
have been extensively  analyzed during the last three decades
\cite{weinpipi,glpl,revpipith}.
After a long period in which hardly any data have been collected,
$\pi\pi$
scattering will  receive in the near future  interesting new input from the
experimental side: i) It is
expected that  forthcoming precise data on $K_{e4}$ decay
 at DA$\Phi$NE and at Brookhaven
 will allow one to determine the phase shift difference
$\delta_{L=0}^{I=0}-\delta_1^1$
near threshold with considerably higher precision \cite{franzini} than
hitherto available \cite{pipiexp}. ii) There are plans to measure
  the lifetime of the $\pi^+\pi^-$--atom in the ground state \cite{nemenov},
and to accurately determine in this manner
the combination $a_0^0-a_0^2$ of the two $S$--wave scattering lengths.
 In order to
confront these
data with  precise theoretical predictions,
it is necessary to go beyond the next--to--leading order term $A_4$
\cite{glpl}. As has
been pointed out in \cite{gchpt} and in \cite{knechtpipi6}, one may eventually
obtain  experimental information on the
size of the quark--antiquark condensate in QCD in this manner.

In Ref. \cite{knechtpipi6}, the part of the amplitude $A_6$
containing branch points -- required by
unitarity -- has been determined. A general crossing symmetric polynomial
of
$O(p^6)$, not fixed by unitarity, has been added by hand.
 In
this letter, we present the analytical result for $A_6$ based on a full
two--loop
calculation in the framework of CHPT. We compare the results of the two
approaches below.

\vspace{.5cm}

\noindent
2. The  expansion (\ref{expansiona}) is most conveniently performed in the
framework of an effective
lagrangian \cite{wein79,glann,hlann}.
Here we consider an expansion around the
chiral limit $m_u=m_d=0$, whereas the  strange quark
mass  is kept at its physical value. We ignore isospin breaking effects and
put $m_u=m_d=\hat{m}$. The
effective lagrangian is expressed in terms of the pion field $U$ and of the
quark mass matrix $\chi$,
           \bea
{\cal L}_{\mbox{\tiny{eff}}}          = {\cal L}_2(U,\chi) + \hbar {\cal
         L}_4(U,\chi) + \hbar^2 {\cal L}_6(U,\chi) + \cdots \co\nonumber
         \eea
where ${\cal L}_n$ contains $m_1$ derivatives of the pion field and
$m_2$
powers of the quark mass matrix, with $m_1+2m_2=n$.
Given ${\cal L}_{\mbox{\tiny{eff}}}$, it is straightforward to expand the
$S$--matrix elements in powers of $\hbar$. This procedure automatically
generates the series (\ref{expansiona}), viz.,
 $A=A_2+\hbar A_4 +\hbar^2 A_6+\ldots.$ The
 leading--order term $A_2$ has been evaluated by Weinberg
\cite{weinpipi}, whereas the next--to--leading order correction $A_4$ was
presented in
\cite{glpl}.
The calculation of $A_6$ requires the
evaluation of two--loop graphs with ${\cal L}_2$, one--loop graphs with one
vertex from  ${\cal L}_4$, and tree graphs generated by
${\cal L}_2 + {\cal L}_4 + {\cal L}_6$. Details of this calculation
will be presented elsewhere. In particular, we refer the reader for the
explicit  expressions of $ {\cal L}_{2,4}$ and of ${\cal L}_6$ to Ref.
\cite{glann} and Ref. \cite{scherer}, respectively.

\vspace{.5cm}

\noindent
3. We use the notation
\bea
&&\langle \pi^d(p_4)\pi^c(p_3)\;\mbox{out}|\pi^a(p_1) \pi^b(p_2)\;\mbox{in}
\rangle =
\langle \pi^d(p_4)\pi^c(p_3)\;\mbox{in}|\pi^a(p_1) \pi^b(p_2)\;\mbox{in}\ra \nn
&&\hspace{3cm}+i
(2\pi)^4\delta^{4}(P_f-P_i)\left\{\delta^{ab}\delta^{cd}A(s,t,u)
+\mbox{permutations} \right\}\co\nonumber
\eea
where $s,t,u$ are the usual Mandelstam variables, expressed in units of the
physical pion mass squared $M_\pi^2$,
\bea
s&=&(p_1+p_2)^2/M_\pi^2 \scs t=(p_3-p_1)^2/M_\pi^2 \scs
u=(p_4-p_1)^2/M_\pi^2\per\nonumber
\eea
Using these dimensionless quantities, the momentum expansion of the amplitude
amounts to a Taylor series in
\bea
{\it x_2}=\frac{M_\pi^2}{F_\pi^2}\co\nonumber
\eea
where $F_\pi$ denotes the physical pion decay constant.
We find
 \bea A(s,t,u)&=& \hspace{.3cm}{\it
x_2}\left[s-1\right]\nn
&&+{\it x_2}^2\left[b_1+b _ 2s + b_3 s^2 +
b_4 ( t - u )^2\right]\nn
&&+{\it x_2}^2\left[F^{(1)}(s) +G^{(1)}(s,t)+G^{(1)}(s,u)\right]\nn
&&+{\it x_2}^3\left[b_5s^3+b_6s(t-u)^2\right]\nn
&&+{\it x_2}^3\left[F^{(2)}(s)+G^{(2)}(s,t)
+G^{(2)}(s,u)\right]\,\!\nn
&&+O({\it x_2}^4)\co
\label{amptot}
 \eea
with
 \begin{eqnarray}
{ F^{(1)}}(\,{s}\,) &=& {\displaystyle \frac {1}{2}}\,
 \bar{J}(\,{s}\,)\,(\,{s}^{2} - 1\,)\co\nn
{ G^{(1)}}(\,{s}, {t}) &=& {\displaystyle \frac {1}{6}}\,
 \bar{J}(\,{t}\,)\,(14\,-\,4\, s\,-\,10\, t\,+\,s\, t\,+\,2\, t^2\,)\co\nn
{ F^{(2)}}(\,{s}\,) &=& \bar{J}(\,{s}\,) \left\{ {\vrule
height0.79em width0em depth0.79em} \right. \! \frac{1}{16 \pi^2}
\left(\! {\displaystyle
\frac {503}{108}}\,{s}^{3} - {\displaystyle \frac {929}{54}}\,{s}
^{2} + {\displaystyle \frac {887}{27}}\,{s} - {\displaystyle
\frac {140}{9}} \right) \nn
 &+&  {b_1}\,(\, 4\,{s}\, - 3) + {b_2}\,(\,
 {s}^{2} + 4\,{s}\, - 4) \nn
 &+& {\displaystyle \frac {{b_3}}{3}}\,\,(\,8\,{s}
^{3} - 21\,{s}^{2} + 48\,{s} - 32\,)
 + {\displaystyle \frac {{b_4}}{3}}\,
(\,16\,{s}^{3} - 71\,{s}^{2} + 112\,{s} - 48\,)
 \! \left. {\vrule height0.79em width0em depth0.79em}
 \right\} \nn
 &+ &\mbox{} {\displaystyle \frac {1}{18}}\,{ K_1}(\,{s}
\,)\, \left\{ \! \! \,20\,{s}^{3} - 119\,{s}^{2} + 210\,{s} - 135 -
{\displaystyle \frac {9}{16}}\,{ \pi}^{2}\,(\,{s} - 4\,)\, \!  \right\}  \nn
 &+ & \mbox{} {\displaystyle \frac {1}{32}}\,{ K_2}(\,{s}\,)\,
 \left\{ \! \,{s}\,{ \pi}^{2} - 24\, \!  \right\}  +
{\displaystyle \frac {1}{9}}\,{ K_3}(\,{s}\,)\,\left\{\,3\,{s}^{2} -
17\,{s} + 9\,\right\}\co \nn
{ G^{(2)}}(\,{s}, {t}\,) &=& \bar{J}(\,{t}\,) \left\{
{\vrule height0.79em width0em depth0.79em} \right. \!
\frac{1}{16 \pi^2}
\left[
{\displaystyle \frac {412}{27}}\! -\! {\displaystyle \frac {s}{54}}
({t}^{2} + 5\,{t} + 159)
\! -\! t \left(\frac{267}{216}{t}^{2} - \frac{727}{108}{t} +
\frac{1571}{108} \right) \right] \nn
 &+&   {b_1}\,(2
 - {t})
+ {\displaystyle \frac {{b_2}}{3}}({t} - 4
)(2\,{t} + {s} - 5)
- {\displaystyle \frac {{b_3}}{6}}
({t} - 4)^{2}(3{t} + 2{s} - 8) \nn
&+& {\displaystyle \frac {{b_4}}{6}}\left(2{s}
(3{t} - 4)({t} - 4) - 32 t + 40t^2 - 11{t}^{3}\,\right)\!
  \left. {\vrule
height0.79em width0em depth0.79em} \right\} \mbox{} \nn
 & +&
{\displaystyle \frac {1}{36}}{ K_1}(\,{t}\,)
\left\{\,174 + 8\,{s} - 10\,{t}^{3}
 + 72\,{t}^{2} - 185\,{t} - {\displaystyle \frac {{\pi}^2}{16}}\,
(\,{t} - 4\,)\,(\,3\,{s}\! -\! 8\,)\, \!
 \right\}  \nn
 &+ & \mbox{} {\displaystyle \frac {1}{9}}\,{ K_2}(\,{t}\,)
\, \left\{ \! \,1 + 4\,{s} + {\displaystyle \frac {{\pi}^2}{64}}
\,{t}\,(\,3\,{s} - 8\,)\
\, \!  \right\}  \nn
 &+& {\displaystyle \frac {1}{9}}\,{ K_3}(\,{t}\,)\left\{
1 + 3{s}{t} - {s} + 3{t}^{2} - 9{t}\right\}
+ {\displaystyle \frac {5}{3}}\,{ K_4}(\,{t}\,)\,\left\{\,4 - 2\,{s} -
{t}\,\right\}\per
\label{amptot1}
\end{eqnarray}
The loop functions $\bar{J}$ and $K_i$ are displayed in appendix A, whereas
the coefficients $b_i$ in the polynomial part
are given in appendix B.
\vspace{.5cm}

\noindent
4. We comment on the structure of the result.
\begin{itemize}
  \item[i)]
 The amplitude $A(s,t,u)$ is expressed in terms of the
external momenta, the physical pion mass, the physical pion decay constant,
and the coefficients $b_1,\ldots ,b_6$.
 To arrive at this result,
 one has to evaluate also $M_\pi$ and $F_\pi$ to two loops\footnote{
 We are indebted to Urs B\"urgi for providing us with the
relevant expression for $M_\pi$ prior to publication \cite{buergi}.}.
Quantum field theory
 leads to  the relations (\ref{constbi}), that
 determine $b_i$ in terms of
\begin{itemize}
\item[--]
 chiral
logarithms $\displaystyle{L=\frac{1}{16\pi^2}\log{\frac{M_\pi^2}{\mu^2}}}\co$
\item[--]
 the low--energy couplings
$l_1^r(\mu),\ldots,l_4^r(\mu)$ from ${\cal L}_4\co$
\item[--]
 the low--energy couplings
$r_1^r(\mu),\ldots,r_6^r(\mu)$ generated by
${\cal L}_6\per$
\end{itemize}
The  scale dependent couplings
$l_i^r$ ($r_i^r$) are needed  to remove the ultraviolet divergences
at  order $p^4$ ($p^6$), and  to generate the most general solution of
the Ward identities at these orders \cite{glann,hlann}.
 The scale $\mu$ -- introduced by the
renormalization procedure --  drops out in the full result (\ref{amptot}).
\item[ii)]
 The contributions
proportional to ${\it x_2}^n$  in (\ref{amptot}) correspond to terms of
$O(p^{2n})$.
 The terms proportional to ${\it
x_2}$ in ${ G^{(2)}}$ and ${ F^{(2)}}$ generate contributions of
  $O(p^8)$ --  these are beyond the accuracy we aim at here.
 In order to keep the formulae
as simple as possible, we nevertheless retain them.
 \item[iii)]
We compare the result  (\ref{amptot})  with the amplitude
given in  Ref.  \cite{knechtpipi6}. Identifying the
low--energy couplings $\alpha,\beta,\lambda_1,\ldots,\lambda_4$ introduced
there with the relevant linear combinations of  $b_1,\ldots,b_6$,
the two expressions agree at $O(p^6)$.
 The $S$--matrix method used in \cite{knechtpipi6} and the
field theory calculation presented here therefore agree as far as the
absorptive part of the amplitude and the general structure of the real
part is concerned.
 Our use of an off--shell method
 provides the additional information Eq. (\ref{constbi}).
Together with an estimate of the
new couplings at  $O(p^6)$, this allows us to make predictions e.g. for the
$S$--waves, see below. In Ref.
\cite{knechtpipi6},  $\alpha$ and $\beta$ are on the other hand treated as
phenomenological parameters that appear in the expressions for the $S$--waves
which remain undetermined.
 Once $\alpha$ and $\beta$ have been pinned down,
 that approach will eventually allow one to determine the size of
the quark--antiquark condensate, and to compare the result
with the Gell--Mann--Oakes--Renner framework \cite{gmor}.
  \end{itemize}
\vspace{.5cm}

\noindent
5. To proceed further, we need to know
 the  values of the coefficients $b_i$. The constants $l_i^r$ that
occur in these have been determined from experimental
data and using the Zweig rule some time ago \cite{glann} (for an update, see
 \cite{bcgke4}).  All these
determinations are faced with
the problem that  the couplings $l_i^r$ are quark mass
independent, whereas the physical quantities, from where
the $l_i^r$ are evaluated,
incorporate quark mass effects. Here we have, for the first time, a means to
pin down these quark mass effects at leading order in an algebraically precise
manner. In order to
achieve this, we need an estimate of the couplings $r_i^r$. This is not an
easy task, and we postpone a complete discussion to a later publication.
Meanwhile, we use for $l_i^r$ the values found in \cite{glann,bcgke4},
\begin{eqnarray*}
\begin{array}{rrrrr}
l_1^r(M_\rho)=&-5.40\cdot 10^{-3}\co&
l_2^r(M_\rho)=&5.67\cdot 10^{-3}\; , \;\\
l_3^r(M_\rho)=&0.82\cdot 10^{-3}\co&
l_4^r(M_\rho)=&5.60\cdot 10^{-3}\;;\; & M_\rho=770 \mbox{~MeV}\per
\end{array}
\end{eqnarray*}
 As for $r_i^r$, we use a by now standard method to obtain an order of
magnitude estimate:
 We incorporate in the above representation the contributions from the
lowest
heavy states and assume that these effects account for the bulk part in
the low--energy couplings at $O(p^6)$. This procedure works
very well at $O(p^4)$ \cite{glann,ressat}. In particular, we  include
the contribution
from vector and scalar exchange, using the couplings presented in
 Ref. \cite{ressat}. In addition to these, kaons and etas also
generate contributions of $O(p^6)$. To estimate those, we have
taken from Ref. \cite{knechtpipi6} (see also Ref. \cite{strassburg}) the
elastic $\pi\pi$ scattering amplitude of $O(p^4)$ evaluated
in the framework of $SU(3)\times SU(3)$
and restricted it to $SU(2)\times SU(2)$ by an expansion in inverse powers
of the strange quark mass.

\vspace{.5cm}
\noindent
6. It is  now straightforward to extract the scattering lengths $a_l^I$ and
slope parameters $b_l^I$.
For the isospin zero $S$--wave, we obtain
\begin{eqnarray*}
{ a^0_0} &=& {\displaystyle \frac {7 {\it x_2} }{32 \pi}}\,
\left\{ {\vrule
height0.93em width0em depth0.93em} \right. \! \! 1 +
{\displaystyle \frac {{\it x_2}}{  \,16\,{ \pi}^{2}\, }}
 \left[ \! \,7 + {\displaystyle \frac {16 \pi^2}{7
}}\,(\,5\,{b_1} + 12\,{b_2} + 48\,{b_3} + 32
\,{b_4}\,)\, \!  \right] \\
& & \; \; \; \; \; \; \; \; \; \; \; \;
 + \left({\displaystyle \frac {{\it x_2}
}{  16{ \pi}^{2}  }}\right)^2 \left[
{\vrule height0.79em width0em depth0.79em} \right. \!
{\displaystyle \frac {7045}{63}}
+ 16\,{ \pi}^{2} \left(
{\vrule height0.79em width0em depth0.79em} \right. \, 10{b_1} + 24{b_2} +
96{b_3}
 + 64{b_4} \\
 & & \; \; \; \; \; \; \; \; \; \; \; \; \; \; \; \; \; \; \; \;
        \; \; \; \; \; \; \; \; \; \; \;
+ {\displaystyle \frac {3072 \pi^2}{7}}
{b_5}\ \! - {\displaystyle
\frac {215}{2016}} \left. {\vrule
height0.79em width0em depth0.79em} \right)   \! \left. {\vrule
height0.79em width0em depth0.79em} \right]
 \! \! \left. {\vrule height0.93em width0em depth0.93em}
 \right\}\co
\end{eqnarray*}
and similarly for the other threshold parameters.
Numerically, we find\footnote{For ease of comparison with earlier calculations,
we use  $F_\pi=93.2$ MeV \cite{pdg90} and $M_\pi=139.57$
MeV.  See also point 9. below. }
 by keeping terms up to and including $O({\it x_2}^3)$
\bea
a_0^0&=&0.217 \; \;  (0.215)\co\nn
a_0^0-a_0^2&=&0.258 \; \; (0.256)\co
\label{numbsl}
\eea
where the numbers in brackets denote the values obtained by putting the
couplings at  $O(p^6)$ to zero at the scale $\mu$= 1 GeV, $r_i^r(1 ~{\rm
GeV})=0$. It is seen that for these threshold parameters the new couplings of
$O(p^6)$ contribute  a negligible amount, if their values are
estimated in the manner described above.
 It is also worth emphasizing
that there is essentially no scale dependence for $0.5 ~{\rm GeV}
\le \mu \le 1 ~{\rm GeV}$. We comment on the theoretical uncertainties
of these and other predictions below. The result (\ref{numbsl})
 should be confronted with the data
\bea
a_0^0&=&0.26\pm 0.05\co\nn
a_0^0-a_0^2&=&0.29\pm 0.04\per\nonumber
\eea
The value for $a_0^0$ is from Ref. \cite{frogg}, whereas we have used the
universal
curve \cite{universal} to express $a_0^2$ through $a_0^0$ and  to obtain
the second relation.
Finally,
we
recall the result for the  one--loop approximation \cite{glpl},
\bea a_0^0&=&  0.201 \;\; (\mbox{one--loop result})\co\nn
a_0^0-a_0^2&=& 0.242 \; \; (\mbox{one--loop result})\per\nonumber
\eea

\vspace{.5cm}

\noindent
7. The chiral expansion of the $S$--wave threshold parameters contains chiral
logarithms \cite{pagels,glpl}.
 At one loop, these are responsible for the bulk part of the correction
to the tree--level result, if the scale in the logarithm is taken at 1 GeV
\cite{glpl}. At two loops, the expansion contains also squares of chiral
logs \cite{colangelo}, e.g.,
\bea
a^0_0 &=& \frac{7 {\it x_2}}{32 \pi } \left\{1 -{\it
x_2} \left[\frac{9}{2}L+\mbox{analytic}\right]\right.\nn
&&\left.\hspace{14.5mm}+{\it x_2}^2
\left[\frac{58}{7}k_1 +\frac{96}{7}k_2+5 k_3+\frac{11}{2} k_4
+\frac{1697}{84}\frac{L}{16\pi^2}+\mbox{analytic}\right]\right.\nn
&&\left.\hspace{9.5mm} +O({\it x_2}^3) \;\;\right\}\co
\label{scattlen2}
\eea
with
\bea
k_i(\mu)=(4l_i^r(\mu)-\gamma_i L)L \; ; \;
\gamma_1=\frac{1}{3}\scs
\gamma_2=\frac{2}{3}\scs
\gamma_3=-\frac{1}{2}\scs
\gamma_4=2\per
\label{kir}
\eea
 The coefficients of the $k_i$'s in the threshold parameters,
in particular in Eq. (\ref{scattlen2}), have been
evaluated earlier in Ref. \cite{colangelo} by means of renormalization group
techniques \cite{wein79}.
Evaluating
the expressions at the scale $\mu=$1 GeV gives
 \bea
a_0^0&=& \overbrace{0.156}^{\mbox{tree}} +\overbrace{0.039 +
0.005}^{\mbox{1~loop}}+
\overbrace{0.013+0.003+0.001}^{\mbox{2~loops}}\;=\;
\overbrace{0.217}^{\mbox{total}}\co\nn
 && \hspace{1.9cm} L \hspace{0.75cm}\mbox{anal.} \hspace{1.1cm} k_i
\hspace{1cm}L
\hspace{0.8cm}\mbox{anal.}\nn\nn
a_0^0-a_0^2&=& 0.201 +0.036 +
0.006+
0.012+0.003+0.001\;=\;
0.258\per\nn
 && \hspace{1.9cm} L \hspace{0.75cm}\mbox{anal.} \hspace{1.1cm} k_i
\hspace{1cm}L
\hspace{0.8cm}\mbox{anal.}\nonumber
 \eea
 We conclude that the nonanalytic terms also
dominate the two--loop corrections in this case.

\begin{figure}[t]
\unitlength1cm
\begin{picture}(2,1) \end{picture}
\epsfysize=8cm
\epsffile{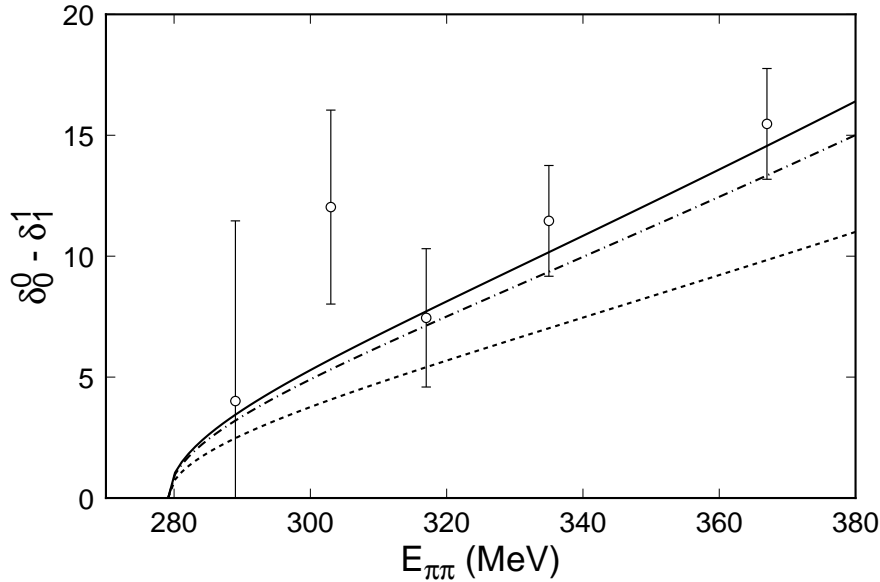}
\caption{The phase shift difference $\delta_0^0-\delta_1^1$ (in degrees) as a
function of the center of mass energy of the two incoming pions. The dotted
(dash--dotted) line displays the tree (one--loop) result,
whereas the solid line denotes the two--loop approximation, evaluated
with $r_i^r(1\;\mbox{GeV})=0$. The data are from Ref.
\protect{\cite{rosselet}}.}
\end{figure}

\vspace{.5cm}

\noindent
8. In Fig. 1, we display the phase shift difference $\delta_0^0-\delta_1^1$
(in degrees) as
a function of the center of mass energy $E_{\pi\pi}=M_\pi\sqrt{s}$. The dotted
(dash--dotted) line stands for the Born term (one--loop result), whereas the
solid line shows the result at two--loop accuracy, evaluated with
$r_i^r(1\;\mbox{GeV})=0.$
It is seen that the two--loop corrections are reasonably small also
considerably
above threshold. Using the couplings $r_i^r$ estimated in the above described
manner increases the two--loop result at 380 MeV by 0.4 degrees.

\vspace{.5cm}

\noindent
9. In summary, we have presented  the analytic expression for the
 elastic
$\pi\pi$ scattering amplitude to two--loop accuracy.
  In order to estimate
 numerically the size of the two--loop contributions, we have determined the
 new couplings
that occur at this order by saturating them with the polynomial contributions
to the amplitude generated by heavy states.
In the
case of the threshold parameters $a_0^0$ and
$a_0^2$, we then find  that i) the new couplings are
numerically negligible,
ii) the bulk part of the correction is due to the presence of chiral
logarithms, and iii) there is no manifestation of a strong enhancement of the
two--loop contributions in this case.

 A more reliable numerical exploitation of the representation (\ref{amptot}),
in particular the evaluation of the remaining uncertainties in the predictions,
 requires additional work: i) As we mentioned above, the quark mass
effects in the determination of $l_i^r$ must be investigated.
 ii) In view
of the accuracy aimed at in future experiments \cite{franzini,nemenov}, isospin
breaking effects cannot be neglected any further. For example, electroweak
radiative corrections must be properly taken into account for extracting
$F_\pi$ \cite{radfpi}. To illustrate,  using $F_\pi=92.4$ MeV
\cite{pdg94} instead of $F_\pi=93.2$ MeV \cite{pdg90} increases
the  values (\ref{numbsl}) for $a_0^0$ $(a_0^0-a_0^2)$ by 0.005 (0.006).
If we use the neutral pion mass instead of $M_{\pi^+}$,
the effect goes in the opposite direction.
 In addition,  real and virtual photon emission in
the scattering process should be investigated.
 We defer these and related issues  to a future publication.

\vspace{1cm}

We thank for useful discussions  or support  B. Anant\-hana\-rayan,
V. Bern\-ard, M. Blatter, U. B\"urgi, P. B\"uttiker, M. Knecht,
U. Mei\ss ner, J. Stern and D. Toublan. We are particularly indebted to H.
Leutwyler for  useful comments that helped us to improve the
original manuscript.   This work is supported in part by  NorFA grant
93.15.078,  by  FWF (Austria), Project No.
P09505--PHY, by Academy of Finland, Project No. 31430,
by HCM,
EEC--Contract
No. CHRX--CT920026 (EURODA$\Phi$NE), and by Schweizerischer Nationalfonds.
Three of us (GE, JG, MS) thank the  Institute for Nuclear
Theory at the University of Washington for hospitality and the DoE for partial
support during the early stage of this work.

   \newcounter{zahler}
\renewcommand{\thesection}{\Alph{zahler}}
\renewcommand{\theequation}{\Alph{zahler}.\arabic{equation}}

\setcounter{zahler}{0}

\appendix

\setcounter{equation}{0}
\addtocounter{zahler}{1}
 \section{Loop functions}
Let
\[
{h}(s)=\frac{1}{N\sqrt{z}}\log
\frac{\sqrt{z}-1}{\sqrt{z}+1} \quad ,\qquad z=1-\frac{4}{s} \; , \;
N=16\pi^2\per
\]
Using matrix notation,  the loop functions used in the text are given by
\begin{eqnarray*}
\left(\begin{array}{l}\bar{J}\\
K_1\\
K_2\\
K_3\\
\end{array}\right)
=
\left(\begin{array}{cccc}
0&0&z&-4N\\
0&z&0&0\\
0&z^2&0&8\\
Nzs^{-1}&0&\pi^2(Ns)^{-1}&\pi^2\\
 \end{array} \right)
 \left(\begin{array}{c}
{h}^3\\
{h}^2\\
{h}\\
\displaystyle{-(2N^2)^{-1}}
\end{array}
\right)\co
\end{eqnarray*}
and
\begin{eqnarray*}
K_4&=&\frac{1}{sz}\left(\frac{1}{2}K_1+\frac{1}{3}K_3+
\frac{1}{N}\bar{J}
+\frac{(\pi^2-6)s}{12N^2}\right)\per\nn
\end{eqnarray*}
The functions $s^{-1}\bar{J}$ and $s^{-1}K_i$ are analytic in the complex
$s$--plane (cut along
the positive real axis for $s \geq 4$), and they vanish
 as $|s|$ tends to infinity. Their real and
imaginary parts are continuous
at $s=4$. The combination $NK_i(s)$ is denoted by $\bar{K}_i(s)$ in
Ref. \cite{knechtpipi6}.

\setcounter{equation}{0}
\addtocounter{zahler}{1}
 \section{The coefficients $b_1,\ldots,b_6$}
The quantities  $b_i$ in Eqs. (\ref{amptot}) and
(\ref{amptot1}) stand for
\begin{eqnarray*}
b_1 &=& 8 l_1^r+2 l_3^r-2 l_4^r+\frac{7}{6} L +
\frac{1}{16{ \pi}^{2}} \frac{13}{18} \\
 &+& {\it x_2}
 \left\{ {\vrule height0.86em width0em depth0.86em} \right. \! \!
\frac{1}{16{ \pi}^{2}} \left[ \frac{56}{9}l_1^r + \frac{80}{9}l_2^r
+15 l_3^r +\frac{26}{9}l_4^r+\frac{47}{108}L-\frac{17}{216}
+  \frac{1}{16{ \pi}^{2}} \frac{3509}{1296} \right]  \\
 & & \; \; \; \; \; \; \;
+ \frac{1}{6}\left[4k_1+ 28k_2-6 k_3+ 13 k_4 \right]
+ \left[32 l_1^r+12 l_3^r - 5 l_4^r \right]
l_4^r - 8 {l_3^r}^2
+ r_1^r
 \left. {\vrule height0.86em width0em depth0.86em} \right\} \co \\ \\
b_2 &=& -8 l_1^r + 2 l_4^r-\frac{2}{3}L
-\frac{1}{16{ \pi}^{2}}\frac{2}{9} \\
&+& {\it x_2}
 \left\{ {\vrule height0.86em width0em depth0.86em} \right. \! \!
\frac{1}{16{ \pi}^{2}}\left[-24 l_1^r-\frac{166}{9}l_2^r-18 l_3^r -
\frac{8}{9} l_4^r -\frac{203}{54}L + \frac{317}{3456}
-\frac{1}{16{ \pi}^{2}} \frac{1789}{432} \right] \\
 & & \; \; \; \; \; \; \; \; \; \; \; \;
-\frac{1}{6}\left[ 54 k_1+62 k_2 + 15 k_3 + 10 k_4 \right]
-\left[ 32 l_1^r+4 l_3^r -5 l_4^r \right] l_4^r +r_2^r
 \! \! \left. {\vrule
height0.86em width0em depth0.86em} \right\} \co \\ \\
b_3 &=& 2 l_1^r + \frac{1}{2}l_2^r -\frac{1}{2}L
-\frac{1}{16{ \pi}^{2}}\frac{7}{12}
+ {\it x_2}
 \left\{ {\vrule height0.86em width0em depth0.86em} \right. \! \!
\frac{1}{16{ \pi}^{2}}\left[
{\vrule height0.86em width0em depth0.86em} \right. \! \!
\frac{178}{9} l_1^r +
\frac{38}{3}l_2^r-\frac{7}{3} l_4^r -\frac{365}{216}L \\
 & & \; \; \; \; \; \; \; \;
- \frac{311}{6912} + \frac{1}{16{ \pi}^{2}}\frac{7063}{864}
\! \! \left. {\vrule height0.86em width0em depth0.86em} \right]
+2 \left[4l_1^r +l_2^r \right]l_4^r
+ \frac{1}{6}\left[38 k_1 + 30 k_2 -3 k_4 \right] + r_3^r
 \! \! \left.
{\vrule height0.86em width0em depth0.86em} \right\}\co \\ \\
b_4 &=& \frac{1}{2} l_2^r -\frac{1}{6}L
-\frac{1}{16{ \pi}^{2}}\frac{5}{36}
+ {\it x_2}
 \left\{ {\vrule height0.86em width0em depth0.86em} \right. \! \!
\frac{1}{16{ \pi}^{2}}\left[
{\vrule height0.86em width0em depth0.86em} \right. \! \!
\frac{10}{9}l_1^r + \frac{4}{9}l_2^r
-\frac{5}{9} l_4^r +\frac{47}{216} L \\
 & & \; \; \; \; \; \; \; \; \; \; \; \;
+\frac{17}{3456}  +\frac{1}{16{ \pi}^{2}} \frac{1655}{2592}
\! \! \left. {\vrule height0.86em width0em depth0.86em} \right]
+ 2 l_2^r l_4^r -\frac{1}{6}\left[k_1+4 k_2 + k_4 \right] + r_4^r
\! \left. {\vrule height0.86em width0em depth0.86em}
 \right\} \co \\ \\
b_5 &=& \frac{1}{16{ \pi}^{2}} \left[-\frac{31}{6} l_1^r -
\frac{145}{36} l_2^r + \frac{625}{288} L + \frac{7}{864}-
\frac{1}{16{ \pi}^{2}} \frac{66029}{20736} \right]\! -\!\frac{21}{16}k_1
\!-\!\frac{107}{96} k_2
+ {\it r_5^r} \co \\ \\
b_6 &=&\frac{1}{16{ \pi}^{2}}
\left[-\frac{7}{18}l_1^r-\frac{35}{36}l_2^r + \frac{257}{864} L
+\frac{1}{432}-\frac{1}{16{ \pi}^{2}}\frac{11375}{20736} \right]
-\frac{5}{48}k_1 -\frac{25}{96}k_2
+ {\it r_6^r}\co
\end{eqnarray*}
\bea
\label{constbi}
\eea
with
$\displaystyle{L=\frac{1}{16\pi^2}\log{\frac{M_\pi^2}{\mu^2}}}$, and where
the $k_i$ are defined in Eq. (\ref{kir}).
We have denoted by  $l_i^r$ ($r_i^r$) the renormalized, quark mass
independent couplings from ${\cal L}_4$ (${\cal L}_6)$, with \cite{glann}
$\displaystyle{\mu\frac{dl_i^r}{d\mu}=-\frac{\gamma_i}{16\pi^2}\per}$
The scale dependences of $r_i^r$ are fixed by the requirement
$\displaystyle{\mu\frac{db_i}{d\mu}=0}$.

\end{document}